\documentclass[prx,twocolumn,longbibliography,superscriptaddress,preprintnumbers]{revtex4-2}
\usepackage[colorlinks,bookmarks=true,citecolor=blue,linkcolor=blue,urlcolor=blue]{hyperref}
\usepackage{dcolumn,graphicx,amsfonts,amsthm,bm,color,appendix,float,tabularx}
\usepackage{braket}
\usepackage[normalem]{ulem}
\usepackage[version=4]{mhchem}
\usepackage{siunitx}
\usepackage{amsmath}
\usepackage{amssymb}
\usepackage{CJKutf8}
\usepackage{multirow}
\usepackage{booktabs}
\usepackage{enumitem}

\usepackage[dvipsnames]{xcolor}
\definecolor{pal0}{rgb}{0.8941, 0.102 , 0.1098}
\definecolor{pal1}{rgb}{0.2157, 0.4941, 0.7216}
\definecolor{pal2}{rgb}{0.302 , 0.6863, 0.2902}
\definecolor{pal3}{rgb}{0.5961, 0.3059, 0.6392}
\definecolor{pal4}{rgb}{1.    , 0.498 , 0.    }
\definecolor{pal5}{rgb}{0.83, 0.69, 0.22}

\usepackage{dsfont}
\newcommand{\A}{\mathcal{A}}

\newcommand{\R}{\mathbb{R}}
\renewcommand{\L}{\mathbb{L}}
\newcommand{\Q}{\mathbb{Q}}

\newcommand{\tr}{\operatorname{\mathsf{tr}}}
\newcommand{\im}{\operatorname{Im}}

\newcommand{\n}[1]{\left| #1 \right|}

\let\caron\v   
\renewcommand{\v}[1]{\boldsymbol{#1}}

\newcommand{\Tr}{\operatorname{Tr}}

\newcommand{\dk}{[d^2 \v{k}]}

\newcommand{\Auc}{\mathcal{A}_{\mathrm{uc}}}
\newcommand{\truc}{\tr_{\mathrm{uc}}}

\renewcommand{\P}{\mathsf{P}}
\renewcommand{\Q}{\mathsf{Q}}
\renewcommand{\H}{\mathsf{H}}
\newcommand{\Con}{\mathsf{C}}
\newcommand{\D}{\mathsf{D}}
\renewcommand{\I}{\mathsf{I}}
\newcommand{\K}{\mathsf{K}}
\renewcommand{\L}{\mathsf{\Lambda}}

\newcommand{\Pk}{\mathsf{P}_{\v{k}}}
\newcommand{\Qk}{\mathsf{Q}_{\v{k}}}
\newcommand{\Hk}{\mathsf{H}_{\v{k}}}

\newcommand{\Pt}{\widetilde{\mathsf{P}}}
\newcommand{\Lt}{\widetilde{\mathsf{\Lambda}}}

\newcommand{\occ}{\mathsf{occ}}
\newcommand{\ucc}{\mathsf{unocc}}

\newcommand{\addcite}[1]{\textsf{\color{red!80!black}[\#]}}

\begin{document}

\title{
Orbital Magnetization within Self-Consistent Hartree-Fock via an Envelope Theorem
}

\author{Daniel E. Parker}
\affiliation{Department of Physics, University of California at San Diego, La Jolla, California 92093, USA}

\begin{abstract}
Orbital magnetization is an elementary thermodynamic response function of a quantum system: the derivative of the grand potential with respect to magnetic field. However, naive perturbation theory is fraught with difficulties, as even infinitesimal magnetic fields shatter any bands into a Hofstadter butterfly spectrum. Even at the self-consistent Hartree-Fock (mean field) level, rigorous formulas for the orbital magnetization have only become available within the last year. This work provides a straightforward derivation of orbital magnetization at the mean field level using a completely real-space approach. The key tools used are the magnetic perturbation theory developed in the 1960s and an envelope theorem, which is a standard result in the calculus of variations. For simplicity, we restrict to insulating states at zero temperature on the infinite plane. We also discuss the natural connections to projected position operators and quantum geometry, as well as practical numerical computations. 
\end{abstract}

\maketitle

\section{Introduction}

Orbital magnetization is notorious for being an extremely difficult linear response property to compute in solid state systems. Its definition is strictly thermodynamic: at $T=0$ in two dimensions
\begin{equation}
	M = - \frac{1}{\mathcal{A}}\frac{d \braket{\hat{\Omega}}}{dB}\bigg|_{B=0},
\end{equation}
where $\mathcal{A}$ is the system area and $\hat{\Omega}(B) = \hat{H}(B) - \mu \hat{N}$ is the zero temperature grand potential at magnetic field $B$. While perturbation theory in magnetic fields was developed in the 1960s~\cite{blount1962bloch,roth1962theory,zak1964magnetic}, even at the single particle level no satisfactory formula for $M$ was available until the ``modern theory of magnetization" in 2005-2007~\cite{thonhauser2005orbital,PhysRevLett.95.137204,ceresoli2006orbital,PhysRevLett.99.197202}. Those works derived the now-standard result under periodic boundary conditions: for single-particle bands $\H \ket{\psi_{\v{k}n}} = \varepsilon_{\v{k}n} \ket{\psi_{\v{k}n}}$ whose wavefunctions have periodic parts $\ket{u_{\v{k}n}} = N^{1/2} e^{-i\v{k}\cdot \hat{\v{r}}} \ket{\psi_{\v{k}n}}$,
\begin{equation}
	\label{eq:single_particle_magnetization}
	M = \frac{e}{\hbar \mathcal{A}} \sum_{\v{k},n} f_{\v{k}n}
\im\braket{\partial_{k_x} u_{\v{k}n}|\H(\v{k}) + \varepsilon_{\v{k}n} - 2\mu|\partial_{k_y}u_{\v{k}n}},
\end{equation}
where $f_{\v{k}n} = f(\varepsilon_{\v{k}n})$ is a Fermi factor. A further decade of study resolved a number of subtle issues about the bulk versus edge nature of the response, culminating in the local ``magnetization marker" formula.

At the mean-field level, a rigorous formula for orbital magnetization only became available in the last year, with the Green's function approach of Kang, Wang, and Vafek~\cite{kang2026orbital}. This work was quickly followed by a flurry of derivations in disparate technical frameworks, including: mean-field finite momentum perturbation theory~\cite{zhu2026orbital,liu2026orbital}; beyond mean-field approaches using finite temperature and local interactions~\cite{chen2026orbital}; and a Luttinger-Ward approach~\cite{ye2026quantum}. Each of these works derives, under varying assumptions, that the magnetization takes the form of Eq.~\eqref{eq:single_particle_magnetization}, but where the single particle bandstructure is directly replaced by the mean-field bandstructure at a self-consistent point. However, most of these derivations adopt an indirect approach, using finite frequency or finite momentum~\cite{huang2026orbital} to compute the static magnetization response.

The goal of this work is to provide a straightforward derivation of the orbital magnetization using physically transparent real-space quantities. Our derivation is performed on the infinite plane, directly in the thermodynamic limit --- suitable for systems with spontaneous symmetry breaking. This allows us to use a uniform perpendicular magnetic field $B$ (instead of one at finite wavevector), and to directly apply the ``magnetic perturbation theory" of the 1960s mentioned above. The workhorse of the derivation is an envelope theorem, which is simple variational calculus. For a mean-field state represented by a single-particle ground state projector $\P$ with mean-field (grand) Hamiltonian $\K =  \H^{\mathrm{HF}}[\P] -\mu \mathsf{N}$, we show that the orbital magnetization takes the form proposed by Resta at the single-particle level~\cite{resta2010electrical}:
\begin{equation}
	\label{eq:mean_field_magnetization}
	M = \frac{e}{2\hbar}\im \tr\big( \; |\K| \; [\D_x,\D_y]  \; \big), 
\end{equation}
where $\tr$ is an intensive trace per unit area, $\D_\mu = [\hat{r}_\mu, \P]$ is a local and bounded avatar of the position operator that can be thought of as an operator-valued Berry connection, and the absolute value of the Hamiltonian is defined as $|\K| = -\P \K \P + \Q \K \Q$ with $\Q = \I - \P$. (Here and below we use \textsf{sans-serif symbols} like $\mathsf{P}$ to denote operators on the single-particle space.) For simplicity, we restrict to insulators at zero temperature.

We make four comments on Eq.~\eqref{eq:mean_field_magnetization}.
(I). All objects in Eq.~\eqref{eq:mean_field_magnetization} are manifestly gauge-invariant, making the formula well-suited for numerical implementation. 
(II).	While Eq.~\eqref{eq:mean_field_magnetization} reduces to Eq.~\eqref{eq:single_particle_magnetization} in the case of discrete translations (shown in App.~\ref{app:momentum_space}), translation symmetry is inessential. Eq.~\eqref{eq:mean_field_magnetization} applies immediately to disordered systems~\cite{schulz2013orbital,bianco2011mapping} (see also~\cite{kang2026orbital}).
(III).		The single-particle version of Eq.~\eqref{eq:mean_field_magnetization} appeared previously in the modern theory of magnetization from Resta and others~\cite{resta2010electrical, schulz2013orbital, bianco2013orbital,marrazzo2016irrelevance,vanderbilt2018berry,resta2020electrical}. Those authors noted it is the average of the local magnetization marker: $M = \mathcal{A}^{-1} \int d\v{r} \; M(\v{r})$. 
(IV).		The commutator $[\D_x,\D_y]$ is an operator-valued Berry curvature over the manifold of projectors, and also appears in the real-space formula for the Chern number~\cite{bellissard1994noncommutative, kitaev2006anyons,bianco2011mapping} 
		\begin{equation}
			\label{eq:kitaev_real_space_Chern}
			C = -2\pi \im \tr \left( \P [\D_x,\D_y] \right),
		\end{equation}
		which corresponds to Hall conductivity $\sigma_{xy} = -e\frac{dn}{dB} = \frac{e^2}{h}C$~\footnote{In our convention, the lowest Landau level in a field $- B \v{e}_z$ with $B>0$ has $C=+1$; TKNN uses $C=-1$ for this instead.}.
 When $\mu$ is in an insulating gap, $P$ is independent of it, so using $| \K| = \K (\I - 2\P)$, $\tr [\D_x, \D_y] = 0$ by cyclicity, and a Maxwell relation immediately yields 
		\begin{equation}
			\frac{dn}{dB} =  \frac{\partial M}{\partial \mu} = \frac{e}{2\hbar}\im \tr\big( (2\P - \I) [\D_x,\D_y] \big) = -\frac{e}{h} C, 
		\end{equation}
		which is the Widom-St\caron{r}eda formula~\cite{widom1982thermodynamic,streda1982theory}.

Our derivation of Eq.~\eqref{eq:mean_field_magnetization} closely follows that of the G\"uttinger-Hellmann-Feynman theorem $\frac{dE}{dB} = \braket{\Psi|\frac{d \hat{H}}{dB}|\Psi}$. However, there is a crucial obstruction: even though $\hat{H}(B)$ is well-defined, $\frac{d \hat{H}}{dB}\big|_{B=0} \sim \hat{ r} \times \hat{v}$ is origin-dependent on the infinite plane. To evade this, we use a switcheroo between the expectation value and the constraint, together with an envelope theorem. The self-consistent Hartree-Fock energy at finite field is a constrained minimization problem
\begin{equation}
	E(B) = \min_{\P} \braket{\hat{\Omega}(B)}_\P \text{ s.t. } \Con[\P] = \P^2 - \P = 0,
\end{equation}
where the Hamiltonian depends on $B$, but the constraint is functionally independent of it. Changing variables to ``twisted" operators $\P \to \Pt$ switches the $B$-dependence from the Hamiltonian to the constraint:
\begin{equation}
	\label{eq:SCHF_B_dependent_constraint}
	E(B) = \min_{\Pt} \braket{\hat{\Omega}(0)}_{\Pt} \text{ s.t. } \Con_B[\Pt] = 0.
\end{equation}
The change of variables is essentially a Peierls phase or Schwinger phase~\cite{schwinger1951gauge}, which was used in this manner as early as the 1960s by Blount~\cite{blount1962bloch}, Roth~\cite{roth1962theory}, and Zak~\cite{zak1964magnetic}, but also more recently in~\cite{kang2026orbital} (see also~\cite{gonze2011density}). After this, $\frac{dE}{dB}$ depends only on the derivative of the constraint $\frac{\partial \Con_B[\Pt]}{\partial B}\big|_{B=0} \propto [\D_x,\D_y]$, which is a well-behaved operator. 

The remainder of this work is organized as follows.
In Section~\ref{sec:mean-field_setup} we provide the mean-field framework for our problem.
Section~\ref{sec:continuum_Hamiltonian} introduces the generic continuum model we work in. 
In Section~\ref{sec:twisted_operators} we introduce the ``twist" change of variables.
In Section~\ref{sec:derivation} we derive Eq.~\eqref{eq:mean_field_magnetization} using the argument outlined above. 
Finally in Section~\ref{sec:discussion} we comment on the physical applicability of our formula, including its relationship to quantum geometry, and its practical numerical computation.

\section{Mean-Field Framework}
\label{sec:mean-field_setup}

We first recall some standard facts about self-consistent Hartree-Fock (see e.g.~\cite{kwan2025mean}), then specialize to the case of interest below. For a grand potential $\hat{\Omega}$, we define the energy density functional $\mathcal{E}[\P] = \braket{\hat{\Omega}}_\P/\mathcal{A}$ (defined explicitly below), whose expectation is with respect to a Slater determinant corresponding to the single-particle density operator $\P = \P^2$. The mean-field ground state is the solution to the constrained optimization problem 
\begin{equation}
	\mathcal{E}^\bullet = \mathcal{E}[\P^\bullet] = \min_{\P} \mathcal{E}[\P]
	\text{ s.t } \Con[\P] = \P^2 - \P = 0.
\end{equation}
whose stationary points satisfy
\begin{equation}
0 = \frac{d \mathcal{E}}{d\P}\Big|_{\P = \P^\bullet}
\end{equation}
under variations that obey the constraint. Here and below the $\bullet$ notation means the quantity is at a particular stationary point. Using Lagrange multipliers, this becomes

\begin{equation}
	\label{eq:SCHF_lagrange_mult}
	\mathcal{E}[\P^\bullet] = \min_\P \max_{\Lambda} \mathcal{L}(\P,\Lambda);\quad
	\mathcal{L}(\P,\Lambda) = \mathcal{E}[\P] + \tr\big[\Lambda \Con[\P]\big],
\end{equation}
where $\tr$ is an intensive trace, defined below. Varying with respect to $\P$ yields
\begin{equation}
	\label{eq:SCHF_EOM}
	\K[\P^\bullet] + \Lambda^\bullet \P^\bullet + \P^\bullet\Lambda^\bullet - \Lambda^\bullet = 0
\end{equation}
where the Hartree-Fock Hamiltonian $\K[\P]$ is a single-particle operator defined by the variation $\delta\mathcal{E}[\P] = \tr\left( \K[\P] \delta \P \right)$. Defining the orthogonal projector $\Q = \I - \P$, the off-diagonal components of Eq.~\eqref{eq:SCHF_EOM} are $\P^\bullet \K[\P^\bullet] \Q^\bullet = 0 = \Q^\bullet \K[\P^\bullet] \P^\bullet$, which implies the self-consistency equation
\begin{equation}
	\label{eq:SCHF_self_consistency}
	\big[ \K[\P^\bullet], \P^\bullet \big] = 0.
\end{equation}
Its solutions are called self-consistent solutions. Meanwhile, diagonal components of Eq.~\eqref{eq:SCHF_EOM}, $\P^\bullet \K[\P^\bullet] \P^\bullet = -\P^\bullet \L^\bullet \P^\bullet$ and $\Q^\bullet \K[\P^\bullet] \Q^\bullet = \Q^\bullet \L^\bullet \Q^\bullet$, the Lagrange multiplier is
\begin{equation}
	\L^\bullet = - \P^\bullet \K[\P^\bullet] \P^\bullet + \Q^\bullet \K[\P^\bullet] \Q^\bullet + \L^\bullet_{\P\Q}
\end{equation}
where $\mathsf{\Lambda}_{\P\Q}^\bullet = \P^\bullet\mathsf{\Lambda}^\bullet\Q^\bullet +\Q^\bullet\mathsf{\Lambda}^\bullet\P^\bullet$ is an underdetermined ``gauge" degree of freedom whose value does not affect the result as it will always appear traced against a block-diagonal quantity. Therefore, 
\begin{equation}
\label{eq:Langrange_multiplier_extrema}
\L^\bullet = -\P^\bullet \K[\P^\bullet] \P^\bullet + \Q^\bullet \K[\P^\bullet] \Q^\bullet = \big|\K[\P^\bullet]\big|,
\end{equation}
where $\n{\K} = -\P \K \P + \Q \K \Q$, which is positive semidefinite. 

\section{Continuum Hamiltonian}
\label{sec:continuum_Hamiltonian}

Consider the infinite plane under a magnetic field $B \v{e}_z = \v{\nabla} \times \v{A}(\v{r})$ minimally coupled to electrons $\hat{\Psi}_a(\v{r})$ with species $a$ and charge $-e$ with $e>0$. Single-particle operators, which can act on the basis $\ket{\v{r},a}$, are denoted by capital \textsf{sans-serif symbols} like $\P$. Whenever possible we work with abstract operators, though a few coordinate expressions are necessary. We work directly in the thermodynamic limit, and use an intensive trace common in the mathematical physics literature~\cite{bellissard1994noncommutative,schulz2013orbital}
\begin{equation}
	\label{eq:intensive_trace}
	\tr[\mathsf{A}] = \lim_{S \to \R^2} \frac{1}{\n{S}}\sum_a \int_S d\v{r} \braket{\v{r},a|\mathsf{A}|\v{r},a}  
\end{equation}
where the limit is taken over a sequence of regions $S$ that expand to fill the entire plane, such as circles of growing radius\footnote{Note that $\tr[\mathsf{U} \mathsf{V}] = \tr[\mathsf{V}\mathsf{U}]$ holds only for sufficiently nice operators, such as Hilbert-Schmidt operators $\tr[\mathsf{U}^\dagger \mathsf{U}] < \infty$, potentially weighted by $\sqrt{\K}$ in some cases below.}.
In the periodic case, Eq.~\eqref{eq:intensive_trace} reduces to the trace over a unit cell: $\tr[\mathsf{A}] = \Auc^{-1} \sum_a \int_{\mathrm{uc}}\braket{\v{r},a|\mathsf{A}|\v{r},a}$.

The electrons are assumed to have a kinetic energy $\mathsf{T}(\hat{\v{p}})$ 
and feel a potential $\mathsf{U}(\hat{\v{r}})$. The grand potential at chemical potential $\mu$ and field $B$ is
\begin{widetext}
\begin{equation}
	\label{eq:grand_potential}
	\hat{\Omega}(B) = \int d\v{r} \, \hat{\Psi}_a(\v{r})^\dagger 
	\left[
		\mathsf{T}_{ab}\big(
			\hat{\v{p}} +e \v{A}(\hat{\v{r}})
		\big)
			+ \mathsf{U}_{ab}(\hat{\v{r}}) - \mu \delta_{ab} 
	\right]
	\hat{\Psi}_b(\v{r})
	+ \frac{1}{2}\int d\v{r}d\v{r}'\,  V_{ab,cd}(\hat{\v{r}}-\hat{\v{r}}') \Psi^\dagger_a(\v{r}) \Psi^\dagger_c(\v{r}') \Psi_d(\v{r}') \Psi_b(\v{r}),
\end{equation}
where repeated indices are summed and $V$ represents density-density interactions. A Slater determinant state is fully specified by the single-particle density matrix  $\P_{ab}(\v{r},\v{r}') = \braket{\hat{\Psi}_b(\v{r}')^\dagger\hat{\Psi}_a(\v{r})}_{\mathrm{SD}}$ where $\P^2 = \P$ . The energy density $\mathcal{E}_B[\P]$ with respect to $\P$, whose coordinate expression we require below, follows from Wick's theorem:
\begin{align}
	\label{eq:HF_energy_density_coordinates}
	&\mathcal{E}_B[\P]
	= \lim_{S\to\infty}
	\frac{1}{\n{S}} \int_S d\v{r} d\v{r}'	\Bigg\{
		[\mathsf{\Omega}_0(B)](\v{r},\v{r}')_{ab}
	\P_{ba}(\v{r}',\v{r})
	+ \frac{1}{2} V_{ab,cd}(\hat{\v{r}}-\hat{\v{r}}') \big[
		\P_{ba}(\v{r},\v{r}) \P_{dc}(\v{r}',\v{r}') - \P_{da}(\v{r}',\v{r}) \P_{bc}(\v{r},\v{r}')
\big]
\Bigg\}
\end{align}
\end{widetext}
with $\mathsf{\Omega}_0(B)
	= \mathsf{T}\big(\hat{\v{p}} +e \v{A}(\hat{\v{r}})\big)
	+ \mathsf{U}(\hat{\v{r}})
	- \mu I$. By Eq.~\eqref{eq:SCHF_lagrange_mult}, self-consistent solutions are stationary points of
\begin{equation}
	\mathcal{L}_B(\P,\Lambda) = \mathcal{E}_B[\P] + \tr(\Lambda \Con_0[\P] ); \quad \Con_0[\P] = \P^2 - \P,
\end{equation}
where the energy part is $B$-dependent but the constraint part is $B$-independent. Whenever there is a differentiable family of stationary solutions $\P^\bullet(B)$ around $B=0$, with energies $\mathcal{E}^\bullet(B)$, the orbital magnetization within self-consistent Hartree-Fock is defined as 
\begin{equation}
	M = - \frac{d \mathcal{E}^\bullet(B)}{dB}\Big|_{B=0}.
\end{equation}

\section{Twisted Operators}
\label{sec:twisted_operators}

To proceed, we introduce ``twisted" operators at finite field, defined by
\begin{equation}
	\Pt(\v{r},\v{r}') = e^{-i \varphi_B(\v{r},\v{r}')} \P(\v{r},\v{r}')
\end{equation}
where
\begin{equation}
	\varphi_B(\v{r},\v{r}') = \frac{e}{\hbar} \int_{\v{r}}^{\v{r}'} \v{A}(\v{r}'') \cdot d\v{r}''
\end{equation}
is the Peierls phase integrated along the straight-line path. This transformation was used as early as the 1960s,
where Zak~\cite{zak1964magnetic} used it to define symmetry-adapted wavefunctions in the classification of irreps of the magnetic translation group, and very similar structures were used in works on magnetic perturbation theory by Roth~\cite{roth1962theory} and Blount~\cite{blount1962bloch}. In the orbital magnetization context, it was applied previously by~\cite{gonze2011density} and~\cite{kang2026orbital}. Physically, this changes the operators from transforming under magnetic translations to transforming under normal translations, which can be thought of as pulling operators back to the $B=0$ Hilbert space. At $B=0$, of course, the transformation is trivial.

Twisting operators changes how multiplication works in the real-space basis. To see this, take operators $\mathsf{U} \mathsf{V} = \mathsf{W}$ or, in coordinates,
\begin{equation}
	\mathsf{W}(\v{r}_1,\v{r}_2) = \int d\v{r}_3 \; \mathsf{U}(\v{r}_1,\v{r}_3)\mathsf{V}(\v{r}_3,\v{r}_2).
\end{equation}
Then
\begin{equation}
	\widetilde{\mathsf{W}}(\v{r}_1,\v{r}_2) = e^{-i\varphi_{12}} \int d\v{r}_3 \; e^{i\varphi_{13}}	\widetilde{\mathsf{U}}(\v{r}_1,\v{r}_3) e^{i\varphi_{32}}	\widetilde{\mathsf{V}}(\v{r}_3,\v{r}_2).
\end{equation}
where $\varphi_{ij} = \varphi_B(\v{r}_i, \v{r}_j) = - \varphi_{ji}$. Multiplication between twisted operators is therefore expressed in terms of a twisted product
\begin{equation}
(\widetilde{\mathsf{U}} \star_B \widetilde{\mathsf{V}})(\v{r}_1,\v{r}_2)
= \int d\v{r}_3 e^{i \Phi_B(\v{r}_1,\v{r}_3,\v{r}_2)}
\widetilde{\mathsf{U}}(\v{r}_1,\v{r}_3) \widetilde{\mathsf{V}}(\v{r}_3,\v{r}_2),
\end{equation}
where $\Phi_B(\v{r}_1,\v{r}_3,\v{r}_2) = \varphi_{13} + \varphi_{32} + \varphi_{21} = \frac{e}{\hbar} \int_{\triangle} B d^2\v{r}$  Here $\triangle$ is the oriented triangle formed by the three points. Since the field is uniform,
\begin{equation}
	\label{eq:triangle_flux}
	\Phi_B(\v{r},\v{r}'',\v{r}') = \frac{eB}{2\hbar} \varepsilon^{\alpha\beta} (r_\alpha - r_\alpha'') (r''_\beta - r_\beta'), 
\end{equation}
where $\alpha,\beta$ are spatial indices.

We note a simple identity required below~\cite{roth1962theory}:
\begin{equation}
	\label{eq:simple_identity}
	\frac{\partial}{\partial B} \left( \mathsf{U} \star_B \mathsf{V} \right)\big|_{B=0}
	= \frac{ie}{2\hbar} \varepsilon^{\alpha\beta} [\hat{r}_\alpha, \mathsf{U}] [\hat{r}_\beta, \mathsf{V}].
\end{equation}
This follows from writing the derivative of the product as
\begin{equation}
	\begin{split}
	&\int d^2\v{r}'' e^{i \Phi_{B}(\v{r},\v{r}'',\v{r}')}
    \frac{i e}{2\hbar} \epsilon^{\alpha \beta}
    (r_{\alpha} - r''_{\alpha})     (r''_{\beta} - r'_{\beta})\\
	&\qquad\times \mathsf{U}(\v{r},\v{r}'') \mathsf{V}(\v{r}'',\v{r}')\Big|_{B=0}
	\end{split}
\end{equation}
and using $(r_\alpha - r''_{\alpha}) \mathsf{U}(\v{r},\v{r}'') = [\hat{r}_{\alpha}, \mathsf{U}](\v{r},\v{r}'')$. 

Putting $\Pt$ into the energy functional Eq.~\eqref{eq:HF_energy_density_coordinates} gives~\cite{kang2026orbital}
\begin{equation}
	\mathcal{E}_B[\P] = \mathcal{E}_{0}[\Pt],
\end{equation}
where $\mathcal{E}_{0}$ is the $B=0$ functional that is, of course, independent of $B$.  Where did all the phases and the vector potential go? 
For the first term, $\mathsf{T}_B = \mathsf{T}(\hat{\v{p}}+e\v{A}(\hat{\v{r}}))$ is shorthand for the kinetic energy conjugated by the vector potential $A$, i.e. 
\begin{equation}
	 \mathsf{T}_B = e^{i \varphi_B} \mathsf{T}(\hat{\v{p}}),
\end{equation}
so $\widetilde{\mathsf{T}}_B = e^{-i \varphi_B} \mathsf{T}_B = \mathsf{T}_{B=0} = \mathsf{T}(\hat{\v{p}})$ and hence $\tr[ \mathsf{T}(\hat{\v{p}}+e\v{A}(\hat{\v{r}}))\P] = \tr[ \mathsf{T}(\hat{\v{p}}) \Pt]$. For the other terms of $\mathcal{E}_B$, the phases vanish: $e^{i \varphi_B(\v{r},\v{r})} = 1$ and  $e^{i \varphi_B(\v{r},\v{r}')}e^{i \varphi_B(\v{r}',\v{r})} = 1$.

The price we must pay for removing the $B$-dependence from the energy is the $B$-independent constraint
\begin{equation}
	\Con_0[\P] = \P^2 - \P = 0
\end{equation}
is replaced by a $B$-dependent one:
\begin{equation}
	\Con_B[\Pt] = \Pt \star_B \Pt - \Pt = 0.
\end{equation}
Self-consistent solutions are now stationary points of
\begin{equation}
	\widetilde{\mathcal{L}}_B(\Pt,\Lt) = \mathcal{E}_0[\Pt] - \tr( \Lt \Con_B[\Pt]) = \mathcal{L}_B(\P,\L),
\end{equation}
which completes the switcheroo. The last equality holds because the phases cancel out inside the trace, which means the energies $\mathcal{E}^\bullet(B)$ are unchanged.

\section{Derivation of Orbital Magnetization}
\label{sec:derivation}
The derivation now proceeds from standard calculus of variations, which we illustrate in a simplified setting for clarity. Suppose we want to extremize a function $f(x)$ subject to a constraint $c(x; \theta) = 0$ that depends on a tuning parameter $\theta$. Defining a Lagrangian $\mathcal{L}(x,\lambda;\theta) = f(x) + \lambda c(x;\theta)$, the stationary conditions
\begin{equation}
	0 = \frac{\partial \mathcal{L}}{\partial x}(x^\bullet, \lambda^\bullet; \theta),
	\qquad 0 = \frac{\partial \mathcal{L}}{\partial \lambda}(x^\bullet, \lambda^\bullet; \theta)
\end{equation}
define a family of extrema $x^\bullet(\theta), \lambda^\bullet(\theta)$. We assume these families are differentiable in $\theta$. If we define $E(\theta) = \mathcal{L}(x^\bullet(\theta), \lambda^\bullet(\theta);\theta)$, then
\begin{equation}
	\frac{dE}{d\theta} =
	\frac{\partial \mathcal{L}}{\partial x}(x^\bullet, \lambda^\bullet; \theta) \frac{d x^\bullet}{d\theta}
	+\frac{\partial \mathcal{L}}{\partial \lambda}(x^\bullet, \lambda^\bullet; \theta) \frac{d \lambda^\bullet}{d\theta}
	+ \frac{\partial \mathcal{L}}{\partial \theta}.
\end{equation}
The first two terms vanish by the stationary condition, and $f(x)$ is independent of $\theta$, leaving
\begin{equation}
	\frac{dE}{d\theta}(\theta) = \left[ \lambda \frac{\partial c}{\partial \theta} \right](x^\bullet(\theta), \lambda^\bullet(\theta); \theta).
\end{equation}
This is known as an envelope theorem. If instead $f$ but not $c$ depended on $\theta$, this would give the G\"uttinger-Hellmann-Feynman theorem.

We may now directly evaluate the orbital magnetization 
\begin{equation}
	M = -\frac{d \mathcal{E}^\bullet(B)}{dB}\Big|_{B=0}
	=   
	-\tr\left[  \Lt\, \frac{\partial \Con_B}{\partial B}  \right]
		\Bigg|_{\substack{\Pt = \Pt^\bullet(0)\\ \Lt = \Lt^\bullet(0)\\B=0}}.
\end{equation}
(Here we have brought the derivative inside the trace, which is permissible because $\Pt$ decays exponentially in $\v{r}-\v{r}'$.) Since the regular and twisted functionals coincide at $B=0$, $\Pt^\bullet(0) = \P^\bullet(0)$ and $\Lt^\bullet(0) = \L^\bullet(0)$, where we have assumed they are differentiable. This is well-justified in the case where $\P$ is an insulator at $B=0$ whose gap should remain open for some range of fields, but is harder to support in metallic states. One place this assumption can fail is where an infinitesimal field can trigger spontaneous symmetry breaking to, e.g., a valley polarized phase. To ease notation, we write $\P^\bullet(0) = \P^\bullet$ and $\L^\bullet(0) = \L^\bullet$. Using the identity Eq.~\eqref{eq:simple_identity} derived above, the $B$-derivative of the twisted constraint at $B=0$ is 
\begin{equation}
	\frac{\partial \Con_B[\Pt]}{\partial B}\Big|_{\substack{\Pt=\P^\bullet\\B=0}} = \frac{ie}{2\hbar} [\D_x,\D_y], 
\end{equation}
with $\D_\mu = [\hat{r}_\mu, \P^\bullet]$. Finally, using $\Lambda^\bullet = | \K[\P^\bullet] |$ from Eq.~\eqref{eq:Langrange_multiplier_extrema}, we have
\begin{equation}
	M = -\frac{ie}{2\hbar} \tr\big( \,| \K[\P^\bullet] |\; [\D_x, \D_y] \, \big).
\end{equation}
Since $\K$ is Hermitian and $\D_\mu$ are anti-Hermitian, the trace is purely imaginary, so $M = -\frac{ie}{2\hbar} (i) \im \tr\big( \,| \K[\P^\bullet] |\; [\D_x, \D_y] \, \big)$, which is Eq.~\eqref{eq:mean_field_magnetization}. 

\section{Discussion}
\label{sec:discussion}

The derivation of M presented here involved three essential objects: the ground state projector $\P$, the (mean field) Hamiltonian $\K$, and the position operator $\hat{r}$. We briefly mention a number of subtle points. First, a fixed size system under periodic boundary conditions only admits integer total flux, making it technically difficult to take the $B\to 0$ limit, motivating our choice of the infinite plane. An alternative is to work with a finite-size droplet, where G\"uttinger-Hellmann-Feynman may be applied directly in the presence of an explicit symmetry-breaking perturbation if necessary. Second, the ``bare" position operator $\hat{r}$ is an unbounded operator on the infinite plane, making it difficult to work with alone. Here, however, it enters only through $\D_\mu = [\hat{r}_\mu, \P]$ or $\D_\mu \P = \Q \hat{r}_\mu \P$. Since $|\P(\v{r}, \v{r}')|$ is ``nearsighted"~\cite{prodan2005nearsightedness},  it decays exponentially in $|\v{r}-\v{r}'|$ whenever it describes an insulator (even a Chern insulator), so $\D_\mu$ is bounded.
This ``tame" avatar of the position operator can then be used without issue. Third, $\P$ is the projector onto \textit{all} filled bands whatsoever, and $\Q$ is the projector onto \textit{all} empty bands. In continuum models with infinitely many bands, $M$ thus involves all ultraviolet physics.
For instance, in a single massive Dirac cone, the closely related self-rotation diverges logarithmically in the ultraviolet cutoff~\cite{li2026contribution}. These issues are not new, but arise already at the single-particle level~\cite{resta2020electrical}.

Much of this derivation should generalize immediately to $T>0$ and metallic states. Indeed, this was already carried out in~\cite{schulz2013orbital} at the single particle level using a similar approach.

Even equipped with Eq.~\eqref{eq:mean_field_magnetization}, computing orbital magnetization in mean field for continuum systems is challenging. This is an especially relevant concern in moir\'e systems, whose orbital magnetization can be measured experimentally to high precision using scanning SQUID-on-tip~\cite{finkler2010self,Vasyukov2013,Grover2022}. Strongly interacting moir\'e systems are often beyond single particle physics, but large swathes of their phase diagrams are well-approximated by self-consistent Hartree-Fock and it is desirous to compute their orbital magnetization~\cite{unconventional_orb_mag,uzan2025hbn,sheekey2026visualizing}. App.~\ref{app:momentum_space} gives a numerically stable formula for $M$ when $\P$ and $\K$ are known in the all-band Hilbert space. In practice, numerics use a finite Hilbert space with an ultraviolet cutoff with perhaps a few hundred bands. Self-consistent Hartree-Fock is then typically performed in a small subset of ``active" bands near the Fermi level, typically two to ten. Crucially, the orbital magnetization cannot be computed from just the active subspace as there are non-zero (dipole) matrix elements connecting all filled bands to all empty ones~\cite{PhysRevLett.133.196501,mao2023diamagnetic,mao2024low}. Ref.~\cite{li2026contribution} recently investigated this issue, finding the computation stabilizes quickly in the total number of bands included in the case of twisted bilayer graphene, but reached no systematic conclusion about the scaling with the size of the active space. Given the $(\varepsilon_{\occ} - \varepsilon_{\ucc})^{-1}$ scaling of the matrix elements (App~\ref{app:momentum_space}), this contribution is expected to decay relatively slowly, but may be small in cases where the active bands are isolated by large band gaps. This issue will be the topic of future work.

While the orbital magnetization itself is not a quantum geometrical response~\cite{resta2017geometrical}, in its derivation we have developed several operator-valued analogues of the Berry curvature. In fact, there are operator-valued counterparts of all standard quantum geometrical quantities. Using $\D_\mu \P = \Q \hat{r}_\mu \P$, one can define an operator-valued quantum geometrical tensor $\hat{\mathcal{Q}}_{\mu\nu} = \P \D_\mu^\dagger \D_\nu \P = \P \hat{r}_\mu \Q \hat{r}_\nu \P$. This is written in terms of the band-theoretical quantum geometric tensor~\cite{souza2000polarization} $\mathcal{Q}(\v{k})$ as $\hat{\mathcal{Q}}_{\mu\nu} = \sum_{\v{k} nm} \ket{\psi_{\v{k}n}} \mathcal{Q}_{\mu\nu}(\v{k})_{nm} \bra{\psi_{\v{k}m}}$. The quantum metric and Berry curvature may be defined from the symmetric and anti-symmetric parts of $\hat{\mathcal{Q}}$ as normal. Further, if one defines a vortexability~\cite{ledwith2023vortexability} operator $\D_{z} \P = [\hat{x} + i \hat{y},\P] \P$, then
\begin{equation}
	(\D_{z} \P)^\dagger (\D_{z} \P) = \P \hat{\bar{z}} \Q \hat{z} \P \ge 0.
\end{equation}
Here $\ge$ means positive semidefinite as an operator. There is equality if, and only if, the vortexability condition $\D_z \P = \Q \hat{z} \P = 0$ holds. None of these definitions make reference to translation symmetry, generalizing the local Chern marker~\cite{bianco2011mapping}.

Finally, we comment that the techniques used here have several natural extensions. One extension is magnetic susceptibility, which was also studied in a recent work using a Green's function approach~\cite{shi2026field}. The same magnetic perturbation theory used here extends naturally to magneto-optical responses and linear or non-linear optical responses of mean-field states.  Beyond electromagnetic $U(1)$, one can also couple to other external gauge fields. For instance, one could couple to strain fields to compute stiffness and other elastic responses relevant to the study of electron crystals~\cite{dong2025phonons}.

\begin{acknowledgements}
	We thank Joel Moore for inspiring us to think about this problem long ago. We acknowledge thank Junkai Dong, Tomohiro Soejima, Eli Zeldov, Oskar Vafek, Jian Kang, Chunli Huang, and especially Minxuan Wang for helpful discussions.
	D.E.P. acknowledges support from NSF CAREER award no. DMR-2542485.
	AI usage: Opus 5.5 (Anthropic) was used to improve the argument and track down minus signs. All text is human written.
\end{acknowledgements}

\bibliographystyle{unsrt}
\bibliography{references_orb_mag}

@article{bianco2013orbital,
  title={Orbital magnetization as a local property},
  author={Bianco, Raffaello and Resta, Raffaele},
  journal={Physical review letters},
  volume={110},
  number={8},
  pages={087202},
  year={2013},
  publisher={APS}
}

@article{thonhauser2005orbital,
  title={Orbital magnetization in periodic insulators},
  author={Thonhauser, Timo and Ceresoli, Davide and Vanderbilt, David and Resta, Raffaele},
  journal={Physical review letters},
  volume={95},
  number={13},
  pages={137205},
  year={2005},
  publisher={APS}
}

@article{resta2010electrical,
  title={Electrical polarization and orbital magnetization: the modern theories},
  author={Resta, Raffaele},
  journal={Journal of Physics: Condensed Matter},
  volume={22},
  number={12},
  pages={123201},
  year={2010}
}

@article{resta2020magnetic,
  title={Magnetic circular dichroism versus orbital magnetization},
  author={Resta, Raffaele},
  journal={Physical Review Research},
  volume={2},
  number={2},
  pages={023139},
  year={2020},
  publisher={APS}
}

@article{kang2026orbital,
  title={Orbital magnetization and magnetic susceptibility of interacting electrons},
  author={Kang, Jian and Wang, Minxuan and Vafek, Oskar},
  journal={Physical review letters},
  volume={137},
  number={10},
  pages={106703},
  year={2026},
  publisher={APS}
}

@article{gonze2011density,
  title={Density-operator theory of orbital magnetic susceptibility in periodic insulators},
  author={Gonze, Xavier and Zwanziger, JW},
  journal={Physical Review B—Condensed Matter and Materials Physics},
  volume={84},
  number={6},
  pages={064445},
  year={2011},
  publisher={APS}
}

@article{schulz2013orbital,
  title={Orbital polarization and magnetization for independent particles in disordered media},
  author={Schulz-Baldes, Hermann and Teufel, Stefan},
  journal={Communications in Mathematical Physics},
  volume={319},
  number={3},
  pages={649--681},
  year={2013},
  publisher={Springer}
}

@article{ye2026quantum,
  title={A quantum many-body approach for orbital magnetism in correlated multiband electron systems},
  author={Ye, Mengxing},
  journal={arXiv preprint arXiv:2601.14372},
  year={2026}
}

@article{zhu2026orbital,
  title={Orbital magnetization and Streda formula of interacting electrons in the mean-field approximation},
  author={Zhu, Jihang and Huang, Chunli},
  journal={Physical Review B},
  volume={113},
  number={20},
  pages={205114},
  year={2026},
  publisher={APS}
}

@article{kitaev2006anyons,
  title={Anyons in an exactly solved model and beyond},
  author={Kitaev, Alexei},
  journal={Annals of Physics},
  volume={321},
  number={1},
  pages={2--111},
  year={2006},
  publisher={Elsevier}
}

@article{chen2026orbital,
  title={Orbital Magnetization of Interacting Electrons},
  author={Chen, Xi and Song, Zhi-Da},
  journal={arXiv preprint arXiv:2602.02478},
  year={2026}
}

@article{liu2026orbital,
  title={Orbital magnetization of correlated states in twisted bilayer transition metal dichalcogenides},
  author={Liu, Xiaoyu and Wang, Chong and Chen, Haoran and Zhang, Xiao-Wei and Cao, Ting and Xiao, Di},
  journal={Physical Review Letters},
  volume={136},
  number={16},
  pages={166606},
  year={2026},
  publisher={APS}
}

@article{streda1982theory,
  title={Theory of quantised Hall conductivity in two dimensions},
  author={Streda, P},
  journal={Journal of Physics C: Solid State Physics},
  volume={15},
  number={22},
  pages={L717--L721},
  year={1982}
}

@article{widom1982thermodynamic,
  title={Thermodynamic derivation of the Hall effect current},
  author={Widom, A},
  journal={Physics Letters A},
  volume={90},
  number={9},
  pages={474},
  year={1982},
  publisher={Elsevier}
}

@article{zak1964magnetic,
  title={Magnetic translation group. II. Irreducible representations},
  author={Zak, J},
  journal={Physical Review},
  volume={134},
  number={6A},
  pages={A1607},
  year={1964},
  publisher={APS}
}

@article{roth1962theory,
  title={Theory of Bloch electrons in a magnetic field},
  author={Roth, Laura M},
  journal={Journal of Physics and Chemistry of Solids},
  volume={23},
  number={5},
  pages={433--446},
  year={1962},
  publisher={Elsevier}
}

@article{schwinger1951gauge,
  title={On gauge invariance and vacuum polarization},
  author={Schwinger, Julian},
  journal={Physical Review},
  volume={82},
  number={5},
  pages={664},
  year={1951},
  publisher={APS}
}

@article{li2026contribution,
  title={Contribution of remote bands to orbital magnetization in twisted bilayer graphene},
  author={Li, Pinzhuo and Jiang, Kun and Wang, Ziqiang and Kang, Jian and Zhang, Yi},
  journal={arXiv preprint arXiv:2603.03702},
  year={2026}
}

@article{mao2023diamagnetic,
  title={Diamagnetic response and phase stiffness for interacting isolated narrow bands},
  author={Mao, Dan and Chowdhury, Debanjan},
  journal={Proceedings of the National Academy of Sciences},
  volume={120},
  number={11},
  pages={e2217816120},
  year={2023},
  publisher={National Academy of Sciences}
}

@article{bianco2011mapping,
  title={Mapping topological order in coordinate space},
  author={Bianco, Raffaello and Resta, Raffaele},
  journal={Physical Review B—Condensed Matter and Materials Physics},
  volume={84},
  number={24},
  pages={241106},
  year={2011},
  publisher={APS}
}

@article{blount1962bloch,
  title={Bloch electrons in a magnetic field},
  author={Blount, EI},
  journal={Physical Review},
  volume={126},
  number={5},
  pages={1636},
  year={1962},
  publisher={APS}
}

@book{vanderbilt2018berry,
  title={Berry phases in electronic structure theory: electric polarization, orbital magnetization and topological insulators},
  author={Vanderbilt, David},
  year={2018},
  publisher={Cambridge University Press}
}

@article{PhysRevLett.99.197202,
  title = {Quantum Theory of Orbital Magnetization and Its Generalization to Interacting Systems},
  author = {Shi, Junren and Vignale, G. and Xiao, Di and Niu, Qian},
  journal = {Phys. Rev. Lett.},
  volume = {99},
  issue = {19},
  pages = {197202},
  numpages = {4},
  year = {2007},
  month = {Nov},
  publisher = {American Physical Society},
  doi = {10.1103/PhysRevLett.99.197202},
  url = {https://link.aps.org/doi/10.1103/PhysRevLett.99.197202}
}

@article{ceresoli2006orbital,
  title={Orbital magnetization in crystalline solids: Multi-band insulators, Chern insulators, and metals},
  author={Ceresoli, Davide and Thonhauser, T and Vanderbilt, David and Resta, Raffaele},
  journal={Physical Review B—Condensed Matter and Materials Physics},
  volume={74},
  number={2},
  pages={024408},
  year={2006},
  publisher={APS}
}

@article{PhysRevLett.95.137204,
  title = {Berry Phase Correction to Electron Density of States in Solids},
  author = {Xiao, Di and Shi, Junren and Niu, Qian},
  journal = {Phys. Rev. Lett.},
  volume = {95},
  issue = {13},
  pages = {137204},
  numpages = {4},
  year = {2005},
  month = {Sep},
  publisher = {American Physical Society},
  doi = {10.1103/PhysRevLett.95.137204},
  url = {https://link.aps.org/doi/10.1103/PhysRevLett.95.137204}
}

@misc{unconventional_orb_mag,
  doi = {10.48550/ARXIV.2506.01485},
  url = {https://arxiv.org/abs/2506.01485},
  author = {Xie,  Jian and Zhang,  Zaizhe and Chen,  Xi and Kwan,  Yves H. and Huo,  Zihao and Herzog-Arbeitman,  Jonah and Guo,  Liangliang and Watanabe,  Kenji and Taniguchi,  Takashi and Liu,  Kaihui and Xie,  X. C. and Bernevig,  B. Andrei and Song,  Zhi-Da and Lu,  Xiaobo},
  title = {Unconventional Orbital Magnetism in Graphene-based Fractional Chern Insulators},
  publisher = {arXiv},
  year = {2025},
  copyright = {Creative Commons Attribution Non Commercial Share Alike 4.0 International}
}

@article{PhysRevLett.133.196501,
  title = {Low-Energy Optical Sum Rule in Moir{\'e} Graphene},
  author = {Mendez-Valderrama, J. F. and Mao, Dan and Chowdhury, Debanjan},
  journal = {Phys. Rev. Lett.},
  volume = {133},
  issue = {19},
  pages = {196501},
  numpages = {10},
  year = {2024},
  month = {Nov},
  publisher = {American Physical Society},
  doi = {10.1103/PhysRevLett.133.196501},
  url = {https://link.aps.org/doi/10.1103/PhysRevLett.133.196501}
}

@article{mao2024low,
  title={Is the low-energy optical absorption in correlated insulators controlled by quantum geometry?},
  author={Mao, Dan and Mendez-Valderrama, Juan Felipe and Chowdhury, Debanjan},
  journal={arXiv preprint arXiv:2410.16352},
  year={2024}
}

@article{resta2017geometrical,
  title={Geometrical meaning of the Drude weight and its relationship to orbital magnetization},
  author={Resta, Raffaele},
  journal={arXiv preprint arXiv:1703.00712},
  year={2017}
}

@incollection{resta2020electrical,
  title={Electrical polarization and orbital magnetization: The position operator tamed},
  author={Resta, Raffaele},
  booktitle={Handbook of Materials Modeling: Methods: Theory and Modeling},
  pages={151--181},
  year={2020},
  publisher={Springer}
}

@article{marrazzo2016irrelevance,
  title={Irrelevance of the Boundary on the Magnetization of Metals},
  author={Marrazzo, Antimo and Resta, Raffaele},
  journal={Physical Review Letters},
  volume={116},
  number={13},
  pages={137201},
  year={2016},
  publisher={APS}
}

@article{finkler2010self,
  title={Self-aligned nanoscale SQUID on a tip},
  author={Finkler, Amit and Segev, Yehonathan and Myasoedov, Yuri and Rappaport, Michael L and Ne’eman, Lior and Vasyukov, Denis and Zeldov, Eli and Huber, Martin E and Martin, Jens and Yacoby, Amir},
  journal={Nano letters},
  volume={10},
  number={3},
  pages={1046--1049},
  year={2010},
  publisher={ACS Publications}
}

@article{Vasyukov2013,
  title = {A scanning superconducting quantum interference device with single electron spin sensitivity},
  volume = {8},
  ISSN = {1748-3395},
  url = {http://dx.doi.org/10.1038/nnano.2013.169},
  DOI = {10.1038/nnano.2013.169},
  number = {9},
  journal = {Nature Nanotechnology},
  publisher = {Springer Science and Business Media LLC},
  author = {Vasyukov,  Denis and Anahory,  Yonathan and Embon,  Lior and Halbertal,  Dorri and Cuppens,  Jo and Neeman,  Lior and Finkler,  Amit and Segev,  Yehonathan and Myasoedov,  Yuri and Rappaport,  Michael L. and Huber,  Martin E. and Zeldov,  Eli},
  year = {2013},
  month = sep,
  pages = {639–644}
}

@article{Grover2022,
  title = {Chern mosaic and Berry-curvature magnetism in magic-angle graphene},
  volume = {18},
  ISSN = {1745-2481},
  url = {http://dx.doi.org/10.1038/s41567-022-01635-7},
  DOI = {10.1038/s41567-022-01635-7},
  number = {8},
  journal = {Nature Physics},
  publisher = {Springer Science and Business Media LLC},
  author = {Grover,  Sameer and Bocarsly,  Matan and Uri,  Aviram and Stepanov,  Petr and Di Battista,  Giorgio and Roy,  Indranil and Xiao,  Jiewen and Meltzer,  Alexander Y. and Myasoedov,  Yuri and Pareek,  Keshav and Watanabe,  Kenji and Taniguchi,  Takashi and Yan,  Binghai and Stern,  Ady and Berg,  Erez and Efetov,  Dmitri K. and Zeldov,  Eli},
  year = {2022},
  month = jun,
  pages = {885–892}
}

@article{kwan2025mean,
  title={Mean-field modeling of moir{\'e} materials: a user's guide with selected applications to twisted bilayer graphene},
  author={Kwan, Yves H and Wang, Ziwei and Wagner, Glenn and Bultinck, Nick and Simon, Steven H and Parameswaran, Siddharth A},
  journal={Advances in Physics},
  volume={74},
  number={1-4},
  pages={11--96},
  year={2025},
  publisher={Taylor \& Francis}
}

@article{ledwith2023vortexability,
  title={Vortexability: A unifying criterion for ideal fractional Chern insulators},
  author={Ledwith, Patrick J and Vishwanath, Ashvin and Parker, Daniel E},
  journal={Physical Review B},
  volume={108},
  number={20},
  pages={205144},
  year={2023},
  publisher={APS}
}

@article{uzan2025hbn,
  title={hBN alignment orientation controls moir$\backslash$'e strength in rhombohedral graphene},
  author={Uzan, Matan and Zhi, Weifeng and Bocarsly, Matan and Dong, Junkai and Dutta, Surajit and Auerbach, Nadav and Kander, Niladri Sekhar and Labendik, Mikhail and Myasoedov, Yuri and Huber, Martin E and others},
  journal={arXiv preprint arXiv:2507.20647},
  year={2025}
}

@article{sheekey2026visualizing,
  title={Visualizing orbital magnetism in electron doped rhombohedral multilayer graphene},
  author={Sheekey, Owen I and Arp, Trevor B and Foutty, Benjamin A and Zhang, Ruoxi and Tan, Tixuan and Holleis, Ludwig FW and Guo, Yi and Kalantre, Sandesh S and Zhang, Canxun and Zakharyan, Mark and others},
  journal={arXiv preprint arXiv:2605.30316},
  year={2026}
}

@article{bellissard1994noncommutative,
  title={The noncommutative geometry of the quantum Hall effect},
  author={Bellissard, Jean and van Elst, Andreas and Schulz-Baldes, Hermann},
  journal={Journal of Mathematical Physics},
  volume={35},
  number={10},
  pages={5373--5451},
  year={1994},
  publisher={American Institute of Physics}
}

@article{souza2000polarization,
  title={Polarization and localization in insulators: Generating function approach},
  author={Souza, Ivo and Wilkens, Tim and Martin, Richard M},
  journal={Physical Review B},
  volume={62},
  number={3},
  pages={1666},
  year={2000},
  publisher={APS}
}

@article{prodan2005nearsightedness,
  title={Nearsightedness of electronic matter},
  author={Prodan, Emil and Kohn, Walter},
  journal={Proceedings of the National Academy of Sciences},
  volume={102},
  number={33},
  pages={11635--11638},
  year={2005},
  publisher={National Academy of Sciences}
}

@article{dong2025phonons,
  title={Phonons in electron crystals with Berry curvature},
  author={Dong, Junkai and Sommer, Ophelia Evelyn and Soejima, Tomohiro and Parker, Daniel E and Vishwanath, Ashvin},
  journal={Proceedings of the National Academy of Sciences},
  volume={122},
  number={38},
  pages={e2515532122},
  year={2025},
  publisher={National Academy of Sciences}
}

@article{shi2026field,
  title={Field Response of the Orbital Moment Probed by Integrated Magnetic Circular Dichroism},
  author={Shi, Qian-Sheng and Wang, Ziqiang and Kang, Jian and Zhang, Yi},
  journal={arXiv preprint arXiv:2610.15684},
  year={2026}
}

@article{huang2026orbital,
  title={Orbital Magnetization from Uniform and Periodic Magnetic Fields},
  author={Huang, Chunli},
  journal={arXiv preprint arXiv:2605.26889},
  year={2026}
}

\clearpage
\newpage

\onecolumngrid
\appendix

\section{Alternative Expression for Orbital Magnetization}

Here we derive an alternative form of Eq.~\eqref{eq:mean_field_magnetization} whose $(\P,\Q)$ block structure is manifest. As in the main text, consider a Hamiltonian $\H$ and $\K = \H - \mu \I$  with ground state projector $\P$ and orthogonal complement $\Q = \I -\P$. Let $\mathsf{D}_{\mu} = [\hat{r}_\mu, \P]$.
Then we have the following identities:
\begin{align}
	\K &= \P \K \P + \Q \K \Q\\
	\P [\mathsf{D}_x, \mathsf{D}_y] \P 
	  &	=  \P \hat{x} \P \hat{y} \P -  \P \hat{y} \P \hat{x} \P
	= - \P \hat{x} \Q \hat{y} \P +  \P \hat{y} \Q \hat{x} \P\\ 
	\Q [\mathsf{D}_x, \mathsf{D}_y] \Q 
	  &=  \Q \hat{x} \Q \hat{y} \Q -  \Q \hat{y} \Q \hat{x} \Q
	= - \Q \hat{x} \P \hat{y} \Q +  \Q \hat{y} \P \hat{x} \Q.	
\end{align}
From Eq.~\eqref{eq:mean_field_magnetization}
\begin{equation}
	M 
	= \frac{e}{2\hbar}\im \tr\left( \; |\K| \;  [\D_x, \D_y] \; \right)
	= \frac{e}{2\hbar}\im \Big[\tr\left( -\P \K \P \;  \P[\D_x, \D_y]\P\right)  + \tr \left(\Q \K \Q \; \Q[\D_x, \D_y]\Q \right)\Big].
\end{equation}
Using the above identity for the commutator in $\P$ block,
\begin{equation}
	\im \tr\left( \P \K \P \;  \P[\D_x, \D_y]\P\right)
	= \im \tr\left( \P \K \P \; \left( - \P \hat{x} \Q \hat{y} \P +  \P \hat{y} \Q \hat{x} \P \right)  \right)
	= \im \tr\left( \Q \hat{x} \P \K \P \hat{y} \Q  - \Q \hat{y} \P \K \P \hat{x} \Q  \right)
	= 2 \im \tr\left( \Q \hat{x} \P \K \P \hat{y} \Q \right).
\end{equation}
To see the last equality, note that if $\hat{W} = \hat{Z} - \hat{Z}^\dagger$, then $\hat{W} - \hat{W}^\dagger = 2\hat{W}$, so
\begin{equation}
	\im \tr[\hat{W}] = \frac{1}{2i} \tr[ \hat{W} -\hat{W}^\dagger ] = \frac{1}{2i} \tr[ 2\hat{W}] =2 \frac{1}{2i} \tr[\hat{Z}-\hat{Z}^\dagger] = 2 \im \tr[\hat{Z}].
\end{equation}
The $\Q$ block is directly analogous so, using anti-symmetry to eliminate the minus sign in the first term,
\begin{equation}
	\label{eq:orb_mag_PQ_blocks}
	M = \frac{e}{\hbar} \im \tr\left( \Q \hat{y} \P \K \P \hat{x} \Q + \P \hat{x} \Q \K \Q \hat{y} \P \right).
\end{equation}
In this form it is clear that only the block off-diagonal components of the position operator ever enter.

The form Eq.~\eqref{eq:orb_mag_PQ_blocks} is not the unique one where the block form is manifest. For instance, other expressions may be derived from applying the identity $[\hat{r}_\mu,\P] \Q = - \P [\hat{r}_\mu, \Q] $, which follows from the fact that $\P\Q=0$ and $[\hat{r}_\mu,\cdot]$ acts as a derivation. 

\section{Connection to the Momentum Space Formula}
\label{app:momentum_space}

We now convert to the momentum space formulation. We first set up notation and recall a few helpful identities from the literature.

Suppose $\H \ket{\psi_{n\v{k}}} = \varepsilon_{n \v{k}} \ket{\psi_{n\v{k}}}$ as usual, and define the periodic parts of the Bloch wavefunctions $\ket{u_{n \v{k}}} = N^{1/2} e^{-i\v{k} \cdot \hat{\v{r}}} \ket{\psi_{n\v{k}}}$, which are normalized on a single unit cell of area $\Auc$. The system size is $\mathcal{A} = N \Auc$. The corresponding Hamiltonian in the $\v{k}$-sector is then $\Hk = e^{-i \v{k} \cdot \hat{\v{r}}} \H  e^{i \v{k} \cdot \hat{\v{r}}}$. 

For any discrete translation-invariant single particle operator $\mathsf{O}$, we similarly define $\mathsf{O}_{\v{k}} = e^{-i \v{k} \cdot \hat{\v{r}}} \mathsf{O}  e^{i \v{k} \cdot \hat{\v{r}}}$, which acts naturally on the Hilbert space of a single unit cell. In particular, the momentum-dependent projectors are
\begin{equation}
	\Pk = \sum_{n} \theta(\mu- \varepsilon_{n\v{k}}) \ket{u_{n\v{k}}} \bra{u_{n \v{k}}}, 
	\qquad \Qk = \mathsf{I}_{\v{k}} - \Pk.
\end{equation}
When $\epsilon_{\v{k} n} < \mu$, then $\Pk \ket{u_{n\v{k}}} = \ket{u_{n\v{k}}}$. Differentiating both sides,
\begin{equation}
	\label{eq:move_derivative_to_ket}
	\Qk \ket{ \partial_\alpha u_{n\v{k}}} = \left( \partial_\alpha \Pk \right) \ket{u_{n\v{k}}}.
\end{equation}

It is convenient to define a trace over the unit cell as
\begin{equation}
	\truc[\mathsf{O}_{\v{k}}] = \sum_{n} \braket{u_{n\v{k}}|\mathsf{O}_{\v{k}}|u_{n\v{k}}} = \int_{\mathrm{uc}} d^2\v{r}\; \braket{\v{r}|\mathsf{O}_{\v{k}}|\v{r}}.
\end{equation}
This trace has a useful identity: if $\hat{O}$ is discrete translation invariant, then~\cite{resta2020magnetic}
\begin{equation}
	\frac{1}{\A} \Tr[\hat{O}] = \frac{1}{\A} \int_{\textrm{sample}} d^2\v{r} \left( \Auc \int \dk \; e^{i \v{k}\cdot \v{r}} \braket{\v{r}|\mathsf{O}_{\v{k}}|\v{r}} e^{-i\v{k} \cdot\v{r}}\right)
	= \frac{\Auc}{\A} \int \dk \;  N \int_{\mathrm{uc}} d^2 \v{r}\braket{\v{r}|\mathsf{O}_{\v{k}}|\v{r}}
	= \int \dk \; \truc[ \mathsf{O}_{\v{k}}].
\end{equation}
Conveniently, this also works for products of operators as $ \frac{1}{\A}\Tr[\hat{O} \hat{O}'] = \int \dk \; \truc[ \mathsf{O}_{\v{k}}  \mathsf{O}'_{\v{k}}]$~\cite{resta2020magnetic}.

We then convert to $k$-space using the following identities (see ~\cite{resta2020magnetic}):
\begin{align}
	\H &= \Auc \int \dk \; e^{i \v{k}\cdot \v{r}} \Hk e^{-i\v{k} \cdot\v{r}},
			&\P &= \Auc \int \dk \; e^{i \v{k}\cdot \v{r}} \Pk e^{-i\v{k} \cdot\v{r}},
			&\D_\alpha &= [\hat{r}_\alpha, \P] = \Auc \int \dk \; e^{i \v{k}\cdot \v{r}} \left( i \partial_\alpha \Pk \right)  e^{-i\v{k} \cdot\v{r}},
\end{align}
where $\dk = d^2\v{k}/(2\pi)^2$. Furthermore $\Q \D_\mu \P = \Q \hat{r}_\mu \P - \Q\P \hat{r}_\mu \P = \Q \hat{r}_\mu \P \to \Qk \left( i \partial_\mu \P_{\v{k}} \right) \Pk$, where $\partial_\mu = \frac{\partial}{\partial k_\mu}$.  

Putting these into Eq.~\eqref{eq:orb_mag_PQ_blocks} yields
\begin{equation}
	M = \frac{e}{\hbar} \int \dk \;
	\im
	\truc\left[
	\Qk \left( \partial_y \Pk \right) \Pk \left( \Hk - \mu \right) \Pk \left( \partial_x \Pk \right) \Qk
	+	\Pk \left( \partial_x \Pk \right) \Qk \left( \Hk - \mu \right) \Qk \left( \partial_y \Pk \right) \Pk 
\right].
\end{equation}
All objects here are manifestly gauge-invariant, making this an excellent practical formula for numerical computations. Ref.~\cite{li2026contribution} used a similar form, whose equivalence follows from $0 = \partial_\mu(\Pk\Qk) = (\partial_\mu \Pk) \Qk + \Pk \left( \partial_\mu \Qk \right)$.

To recover the standard formula Eq.~\eqref{eq:single_particle_magnetization} takes a few more steps. By cyclicity of the trace,
\begin{align}
	M &= \frac{e}{\hbar} \int \dk \;
	\im
	\truc\left[
	\Pk \left( \partial_x \Pk \right) \Qk \left( \Hk - \mu \right) \Qk \left( \partial_y \Pk \right) \Pk 
	+ 
	\left( \Hk - \mu \right) \Pk \left( \partial_x \Pk \right) \Qk \left( \partial_y \Pk \right) \Pk
\right]\\
	  &= \frac{e}{\hbar} \sum_{n \in \occ} \int \dk \; 
	  \im \Big\{
	  \Braket{u_{n\v{k}}|
		\left( \partial_x \Pk \right) \Qk \left( \Hk - \mu \right) \Qk \left( \partial_y \Pk \right)
	  |u_{n\v{k}}}
	  + (\varepsilon_{n\v{k}} - \mu)
	  \Braket{u_{n\v{k}}|
		\left( \partial_x \Pk \right) \Qk \left( \partial_y \Pk \right) 
	  |u_{n\v{k}}}
  \Big\}.
\end{align}
Applying Eq.~\eqref{eq:move_derivative_to_ket} then gives
\begin{equation}
	\label{eq:M_du_form_stable}
	M = \frac{e}{\hbar} \sum_{n \in \occ} \int \dk \; \im
	\braket{\partial_x u_{n\v{k}}| \Qk \left(\Hk + \varepsilon_{n \v{k}} - 2\mu\right) \Qk |\partial_y u_{n\v{k}}}.
\end{equation}
This has the form of Eq.~\eqref{eq:single_particle_magnetization} except for the extraneous $\Qk$'s. In fact, these may be eliminated. To see this, start with 
\begin{equation}
	\frac{e}{\hbar} \sum_{n \in \occ} \int \dk \; \im
	\braket{\partial_x u_{n\v{k}}| (\Qk+\Pk) \left(\Hk + \varepsilon_{n \v{k}} - 2\mu\right) (\Qk+\Pk) |\partial_y u_{n\v{k}}}.
\end{equation}
Since $\Qk\Pk = 0$ and $\Qk \H_k \Pk=0$, only the diagonal blocks are non-zero. The $(\Pk,\Pk)$ (occupied-occupied) contribution is separately zero since the $(n,m)$ and $(m,n)$ contributions to
\begin{equation*}
	\sum_{n,m\in\occ}(\varepsilon_{n\v{k}}+\varepsilon_{m\v{k}}-2\mu)\braket{\partial_x u_{n\v{k}}|u_{m\v{k}}}\braket{u_{m\v{k}}|\partial_y u_{n\v{k}}}
\end{equation*}
are complex conjugates whose imaginary part vanishes. We have thus recovered Eq.~\eqref{eq:single_particle_magnetization}. We comment that dropping the $\Qk$ terms is not advisable because the occupied-occupied contribution often cannot be computed stably in practice. 

For completeness, we also demonstrate the equivalence to the expression for $M$ in terms of matrix elements of the velocity operator. The (Hartree-Fock) velocity operator in band theory has the form
\begin{equation}
	\mathsf{v}_{\v{k}} = \frac{1}{\hbar} \v{\nabla}_{\v{k}} \Hk,
\end{equation}
which is a gauge-invariant quantity that may be evaluated by finite differences. Its matrix elements obey
\begin{equation}
	\braket{u_m|\partial_\alpha u_n} = -\hbar \frac{\mathsf{v}^\alpha_{mn}}{\varepsilon_m-\varepsilon_n} \quad (n \neq m).
\end{equation}
Putting this into Eq.~\eqref{eq:M_du_form_stable} yields 
\begin{equation}
		\label{eq:orbital_magnetization_velocity}
		M = e \hbar    \sum_{\substack{n \in \occ\\m \in \ucc}} \int \dk \; \frac{(\varepsilon_{m \v{k}} + \varepsilon_{n \v{k}} - 2\mu)}{(\varepsilon_{n\v{k}} - \varepsilon_{m\v{k}})^2}  \im \mathsf{v}^x_{nm} \mathsf{v}^y_{mn}.
\end{equation}

\twocolumngrid

\end{document}